\documentclass[11pt]{article}
\usepackage{epsfig,enumerate,verbatim}
\usepackage{amsmath}
\usepackage{amsfonts}
\usepackage{amssymb}
\usepackage{amstext}
\usepackage{amsthm}
\usepackage{undertilde} 
\usepackage{dsfont}
\usepackage{graphicx}
\usepackage{mathrsfs}
\usepackage{cite}
\begin{document}        
\title{\bf{Generalized Relativistic Wave Equations with Intrinsic Maximum Momentum}}
\author{Chee Leong Ching\footnote{Email: phyccl@nus.edu.sg} and Wei Khim Ng\footnote{Email: phynwk@nus.edu.sg}}

\date{August 2013 V2}

\maketitle

\begin{center}
{Department of Physics,\\}
{National University of Singapore,\\}
{Kent Ridge,\\}
{Singapore.}
\end{center}

\begin{abstract}
\noindent We examine the nonperturbative effect of instrinsic maximum momentum on the relativistic wave equations. Using the momentum representation, we obtain the exact eigen-energies and wavefunctions of one-dimensional Klein-Gordon and Dirac equation with linear confining potentials, and Dirac oscillator. Similar to the undeformed case, bound state solutions are only possible when the strength of scalar potential are stronger than vector potential. The energy spectrum of the systems studied are bounded from above, whereby classical characteristics are observed in the uncertainties of position and momentum operators. Also, there is a truncation in the maximum number of bound states that is allowed. Some of these quantum-gravitational features may have relevant applications in quarkonium confinement and quantum gravity phenomenology.
\end{abstract}


\section{Introduction}

Linear confining potentials in the relativistic regime are important quantum mechanical models to examine the phenomenology of quarkonium and their confinement properties. It was well studied that when the linear potential is introduced as a Lorentz scalar in the relativistic wave equation (RWE) such as Klein-Gordon (KG) or Dirac equation, quantum confinement is possible \cite{suRK1, suRK2}. This is because a scalar potential in RWE is equivalent to a position dependent of the rest mass similar to the MIT bag model of quark confinement \cite{alberto}.

If the potential is introduced as a Lorentz vector, no bound state solutions are possible due to the existence of the Klein paradox \cite{gunion}. In this case only tunneling solutions are possible. For the mixtual of both scalar-like and vector-like potential, bound states can only exist if the strength of scalar-like potential is stronger than the vector-like potential \cite{suRK3}. For the critical case when the strength of the both potentials are exactly equal, confinement is impossible and only scattering solution arises \cite{suRK2}.  

Dirac oscillator is the dirac equation in which the momentum is replaced by the non-minimal coupling $\vec{P} \rightarrow \vec{P} - i m \omega \beta \vec{r}$, where $\vec{r}$ is the position operator, $m$ is the particle mass, $\beta$ is the usual Dirac matrix and $\omega$ is the frequency of the oscillator. It has been used to model quark confinement in strong force. The non-relativistic limit reproduces the usual harmonic oscillator with very strong spin-orbit coupling. These properties are valid under usual Heisenberg commutation relation.   

Modified commutation relations (MCR's) have been extensively studied as an effective means of encoding potential gravitational or stringy effects \cite{kempf, snyder, laynam, brau, hassan, sabine}. It has also been suggested that the consequent deformations of quantum mechanical spectra might be detectable in future low-energy experiments \cite{laynam, brau, optics}.

While most of the studied MCR's incorporate a minimum position uncertainty and usually leads to the concept of minimal length \cite{kempf, laynam, brau} which generally known as the main prediction by generalized uncertainty principle (GUP), there are others that exhibit a maximum momentum cutoff \cite{dsr, das, das-rel, mig,jiz,cutoff, pedrammax, spectra} as in Doubly Special Relativity (DSR). We have investigated large classes of deformed quantum mechanics of the latter type and used it to study several quantum mechanical systems such as deformed harmonic oscillator (DHO) \cite{spectra}, a particle in potential well/barrier and their bound/scattering states \cite{potwell} as well as generalized coherent states with maximum momentum \cite{coherent}. 

In one-dimension, the relevant modified quantum algebra is
\begin{equation}
[X,P] = i\hbar\ f(P)  \, , \label{mod}
\end{equation}
with $f(P) =1$ in usual Heisenberg's quantum mechanics. Intrinsic maximum momentum arises when $f(P)$ has a singularity \cite{pedrammax, spectra} or a zero at some $P=P_0$ \cite{dsr,das, das-rel, mig,jiz,cutoff,spectra}. One effect of maximum momentum is that the spectrum of bound states terminates at finite energy even for potentials like the harmonic oscillator \cite{spectra}; this is in contrast to MCR's which exhibit instead a minimum position uncertainty \cite{kempf,laynam, brau}. 

In Ref.\cite{spectra} we focussed on the class of MCR's defined by characteristic function 
\begin{equation}
f(P) = 1- 2\alpha P + q\alpha^2 P^2 \label{class}
\end{equation} 
where $\alpha>0$ and $q$ are real parameters. Indeed, $\alpha$ is the deformed parameter with dimension of inverse momentum, $\alpha=\frac{\alpha_{0}}{M_{pl}c}=\frac{\alpha_{0}L_{Pl}}{\hbar}$ such that $M_{pl}\approx 10^{19} GeV/c^2$ is the Planck mass and $L_{pl}\approx 10^{-35}m$ is the Planck length. The dimensionless parameter $\alpha_{0}$ is usually assumed to be of order unity, which implies the $\alpha$-dependent terms are only important in the Planck regime, i.e. when energy and momentum are comparable to Planck energy (momentum) and the size is of order Planck length. This form of MCR leads to DSR-inspired modified uncertainty principle (MUP) which is consistent with string-theory GUP and results from black hole physics. 

For $q \le 1$, eq.\eqref{class} has a zero and hence (in momentum space) eq.\eqref{mod} implies an intrinsic maximum momentum. For $q>1$ there is no intrinsic maximum momentum but rather a minimum position uncertainty. For simplicity, we consider only the case $q=0$ and the generic result should be similar for all $q\leq 1$.

In Sect.(2) we review the main relations of quantum mechanics based on MCR's and discuss the ideas of minimal length and intrinsic maximum momentum cutoff. In Sect.(3), modified dispersion relation is considered up to leading order correction. Superluminal propagation of photons is possible by MCR in the Planck regime which consistents with the varying speed of light phenomena. In Sect.(4) we study the one-dimensional free (and with both scalar/vector-like potential) KG equation. Using momentum and position representation interchangeably, we solve the bound state energy spectrum and corresponded wavefunctions. We show that, for spinless particle satisfies KG wave equation, the energy spectrum is bounded from above and hence the number of the bound state solutions is finite. This is the generic result of MCR's with intrinsic maximum momentum cutoff as in their non-relativistic counterpart \cite{spectra,potwell,coherent}. In Sect.(5) and (6), with the same token we consider the Dirac equation with scalar/vector linear potential and Dirac oscillator respectively. The bound state energy spectrum, wavefunctionas and number of state are determined. Also, we compute the MUP for Dirac oscillator and show that the non-relativistic limit reproduces same feature as DHO. Finally we conclude in Sect.(7).        

\section{Quantum Mechanics with Modified Commutation Relations (MCR)}

The full set of three-dimensional deformed commutation relation is given by 
\begin{eqnarray}
\left[X_{i}, X_{j}\right]&=&\left[P_{i}, P_{j}\right]=0\hspace{0.3cm}\text{for all $(i,j)$}\\
\left[X_{i}, P_{j}\right]_{r}&\approx&i\hbar\left[\delta_{ij}-\alpha\Bigl(\delta_{ij}P+\frac{P_{i}P_{j}}{P}\Bigr)+\alpha^2(\delta_{ij}r P^2+(2r+1)P_{i}P_{j})\right]\label{class1}
\end{eqnarray}
where $\alpha$ and $r$ are parameters to specify the MCR. For $r=0$, Eq.\eqref{class1} corresponds to the one-parameter deformed commutator which is at most quadratic in momenta and which satisfies the Jacobi identity exactly. For one-dimension, \eqref{class1} reduces to \eqref{mod} and \eqref{class} with $q=3r+1$. Note that the coordinate space is commutative.

Next, consider an explicit representation of the position and momentum operators obeying relation \eqref{mod} and \eqref{class} in momentum space
\begin{eqnarray}
P\ \phi(p)&=&p\ \phi(p)\label{rep.P}\\
X\ \phi(p)&=& i\hbar f(p)\frac{\partial}{\partial p} \phi(p)= i\hbar \Bigl(1- 2\alpha p + q\alpha^2 p^2\Bigr)\frac{\partial}{\partial p} \phi(p). \label{rep.X}
\end{eqnarray}
It is clear that for real momentum $p$ the polynomial $f(p)$ has roots at $(1\pm \sqrt{1-q} )/ (\alpha q)$. Thus for $q>1$ the roots are away from the real line and \eqref{mod} and \eqref{rep.X} are well defined. However for $q=1$ there is a doubly degenerate real root and a momentum cutoff  $P< 1/(q\alpha)$ is required. For $q<1$ there are two real roots except at the point $q=0$ where there is only a single real root at $1/(2\alpha)$. The position and nature of these roots determine qualitatively the different features of the deformed spectrum \cite{spectra}.

We obtain the resulting uncertainty relation from the sub-class of MCR from \eqref{class} 
\begin{equation}
\Delta X \Delta P \ \ge \ {1 \over 2} \langle[X,P]\rangle \ \ge \ {\hbar \over 2} ( 1 -  2\alpha \langle P \rangle +  q\alpha^2 \langle P^2 \rangle ) \ .\label{mcr}
\end{equation}
This can be written as the MCR-inspired MUP
\begin{equation}
{2 \Delta X \over \hbar} \ge \frac{(1 -\alpha \langle P \rangle)^2}{\Delta P} + \frac{(q-1)\alpha^2 \langle P \rangle^2}{\Delta P} +  q\alpha^2 (\Delta P) \label{mcr1}
\end{equation}
for all real $q$ parameter. Particularly, for $(q-1)>0 $ the right side of (\ref{mcr1}) is positive and increases when $\Delta P$ tends to zero or infinity.  
Thus there exists a sub-class of the exactly realised MCR's for which $\Delta X$ has a nonzero minimum; such a minimum position uncertainty is usually associated with some fundamental length scale \cite{sabine} and originally motivated by the search for deformed Heisenberg algebras that would realise such scenarios \cite{kempf,cutoff}. Simply on dimensional grounds, $(\Delta X)_{min} \sim O(\hbar\alpha)$. 

$q=1$ in \eqref{class} is an exact MCR with maximum momentum cutoff $P_{max}=1/\alpha$ \cite{spectra}. For the truncated versions of the MCR's displayed in (\ref{class}), $q>1$ seems to allow for a minimum position uncertainty but no intrinsic momentum cutoff while for $q \le 1$  we have a maximum momentum cutoff but apparently no minimum position uncertainty. Beside the minimal length which is predicted by quantum gravity theory, an intrinsic maximum momentum of $O(1/\alpha)$ which is favoured by deformed special relativity \cite{dsr} can be realised with MCR's in (\ref{class}) with certain range of $q$ values.

It is worthwhile to note that the MCR-inspired MUP \eqref{mcr1} also exhibits a peculiar form of the UV/IR mixing that may allow short distance physics to be probed by large scale physics and vice versa. From \eqref{mcr1}, when $\Delta P$ is large (UV), $\Delta X$ is proportional to $\Delta P$ (for $q\neq 0$). Thus $\Delta X$ is also large (IR). This scenario is expected from most of the quantum gravity phenomenology such as AdS/CFT correspondence and noncommutative field theory. 

Since the MCR in \eqref{class} exhibits minimal length for certain range of $q$ values, i.e. $q>1$, we do not have the notion of localization in the position space for all $q$ values. By construction, there is no minimal momentum in MCR's. Thus, momentum space becomes more convenient for the analysis of any eigenvalue problem \footnote{We will concentrate on $q=0$ case in which there is maximum momentum cutoff but without minimal position uncertainty. In fact, for $q\leq 1$ we have the flexibility to use either position or momentum space.}. Hermicity of $X$ operator \eqref{rep.X} requires the weight
function $1/f(p)$ in the inner product:
\begin{eqnarray}
\langle\phi|\psi\rangle&=&\int_{-\infty}^{p_{max}}\frac{dp}{(1-2\alpha p+q\alpha^2 p^2)}\phi^*(p)\psi(p).
\end{eqnarray}
Note that the upper integration limit has to be set to $p=p_{max}$ for the case $q\leq 1$ in order for the convergence of the integral involved. For $q>1$, $p_{max}\rightarrow \infty$.   

\section{MCR Dispersion relation and Varying Speed of Light}

For $q\leq 1$, we consider the position representation since there is no minimum position uncertainty \cite{potwell}. The exact representation in position space for $q=1$ is
\begin{eqnarray}
X^{i}&=&x^{i}\\
P^{i}&=&\frac{p^{i}}{1+\alpha |p|}
\end{eqnarray} 
while for $q=0$ 
\begin{eqnarray}
P^{i}&=&\frac{1}{2\alpha}(1-e^{-2\alpha p^{i}}). \label{exactP}
\end{eqnarray} 
Here $(x^{i},p^{i})$ are usual low energy canonical pairs that satisfy the Heisenberg algebra, $[x^{i},p^{j}]=i\hbar\delta^{ij}$ and $|p|=\sqrt{g_{ij} p^{i} p^{j}}$ is the magnitude of the 3-momentum. We denote $P^{\mu}$ as the four momentum\footnote{Greek indices run from $(\mu,\nu= 0,1,2,3)$ whereas italic indices run from $(i,j=1,2,3)$.}. Thus for $q=1$ we have   
\begin{eqnarray}
X^{\mu}&=&(x^0, x^{i})\\
P^{\mu}&=&\bigl(P^{0} = E/c,\ P^{i} = p^{i}\ (1-\alpha |p| + \alpha^2 |p|^2)\bigr) \label{defP}
\end{eqnarray} 
while for $q=0$ 
\begin{eqnarray}
P^{\mu}&=&\bigl(P^{0} = E/c,\ P^{i} =p^{i}\ (1-\alpha |p| + \frac{3}{2}\alpha^2 |p|^2)\bigr)
\end{eqnarray} 
where perturbatively we only keep terms up to $O(\alpha^2)$ \cite{bibhas}. Consider $q=1$ and denote the generic background metric as $g_{\mu\nu}$, the norm of $P^{\mu}$ is
\begin{eqnarray}
P^{\mu}P_{\mu} &=& g_{\mu\nu}P^{\mu}P^{\nu} = g_{00}\bigl(P^{0}\bigr)^2 +g_{ij} P^{i}P^{j}\nonumber\\
&=& g_{00}\frac{E^2}{c^2} + g_{ij} p^{i} p^{j}\ \bigl(1-\alpha |p| + \alpha^2 |p|^2\bigr)^2\nonumber\\
&=& g_{00}\frac{E^2}{c^2} + |p|^2 + |p|^2 \bigl(-2\alpha |p| + 3 \alpha^2 |p|^2\bigr). 
\end{eqnarray}

Since the undeformed dispersion relation is $p^{\mu}p_{\mu}=g_{00} E^2/c^2 + |p|^2$ where $p^{\mu}p_{\mu}$ is the Casimir invariant of the usual Lorentz group given by the term $-m^2 c^2$, we obtain the modified dispersion relation in terms of low energy momentum as
\begin{eqnarray}
P^{\mu}P_{\mu} &=& -m^2 c^2 + |p|^2 \bigl(-2\alpha |p| + 3 \alpha^2 |p|^2\bigr). \label{disp}
\end{eqnarray}

One can invert \eqref{defP} to express the undeformed momentum $p^{i}$ in terms of deformed $P^{i}$, i.e. $p^{i}= P^{i}(1+ \alpha \hbar |P|)$ where $|P|=\sqrt{g_{ij} P^{i}P^{j}}$. For simplicity, we keep up to leading order correction $O(\alpha)$. Thus, \eqref{disp} takes the form of 
\begin{eqnarray}
P^{\mu}P_{\mu} &=& -m^2 c^2 -2 \alpha |P|^3=-m_{\text{eff}}^2\ c^2\ ,\label{disp1}
\end{eqnarray}
where we define the modified effective mass due to MCR as $m_{\text{eff}}:=\sqrt{m^2 + 2 \alpha |P|^3/c^2}$. This correction can be viewed as the correction due to Planck length physics. Making used of $P^{\mu}P_{\mu}=g_{00} (P^{0})^2 + |P|^2$, Eq. \eqref{disp1} can be written as
\begin{eqnarray}
(P^{0})^2 &=& -\frac{1}{g_{00}}\left[m^2 c^2 + |P|^2 (1 + 2 \alpha |P|)\right]\ ,\label{disp2}
\end{eqnarray}
and thus the energy of a particle in the gravitational field $g_{\mu\nu}$ \cite{padm} with MCR is
\begin{eqnarray}
E^2 &=& (-g_{00} c P^{0})^2= - g_{00} c^2 \left[m^2 c^2 + |P|^2 (1 + 2 \alpha |P|)\right]\label{disp3}\\
&=& E_{0}^2 - 2 g_{00} c^2 \alpha |P|^3.\nonumber
\end{eqnarray}    
Here $E_{0}^2=- g_{00} c^2 (m^2 c^2 + |P|^2)$ is the usual dispersion relation. If we consider Minskowski spacetime $g_{00}=-1$ and set $m=0$, the photon's group velocity is define by $v_{p}=\partial E/\partial |P|$. From \eqref{disp3} one can solve $|P|$ in terms of energy $E$ by iteration,     
\begin{eqnarray}
E^2 &=& c^2 |P|^2 \bigl(1 + 2 \alpha |P|\bigr)\nonumber\\
\Rightarrow |P|&=& \pm \frac{E}{c}\left(1 \pm \frac{\alpha E}{c}\right).
\end{eqnarray}    
Hence up to $O(\alpha)$, the modified photon velocity is , 
\begin{eqnarray}
v_{p}&=&\partial E/\partial |P| = \pm c (1 + 2 \alpha |P|)\nonumber\\
&\approx& c \left(\pm 1 + \frac{2\alpha E}{c}\right)=c \left(1 + \frac{2\alpha E}{c}\right)
\end{eqnarray} 
where we choose ``+" sign by the consistency condition, i.e. $v_{g}= c$ as $\alpha\rightarrow 0$. Thus we have energy dependent velocity of light at the Planck length regime. This superluminal photon propagation, i.e. $v_{g}> c$ is consistent with Varying Speed of Light (VSL) phenomena \cite{petit, moffat, magueijo}.   

\section{MCR and the Real Klein-Gordon equation}

In relativistic quantum mechanics, the equation of motion of a quantum scalar or pseudoscalar field which describes the spinless particles is given by the Klein-Gordon (KG) equation. In (1+1)-dimensions, it is given by \cite{greiner1}
\begin{eqnarray}
c^2 p_{x}^2 \phi(x,t)=(E^2-m^2c^4)\phi(x,t). \label{kge1}
\end{eqnarray}
We recognise \eqref{kge1} as the classical wave equation including the mass term $m^2c^4$. If we choose the representation $p_{x}=-i \hbar\partial_{x}$ and $E=i\hbar \partial_{t}$, we end up with the free KG equation
\begin{eqnarray}
\Bigl(\frac{\partial^2}{c^2\partial t^2}-\frac{\partial^2}{\partial x^2}+\frac{m^2c^2}{\hbar^2}\Bigr)\phi(x,t)=0\label{kge2}
\end{eqnarray}
with the free solution $\phi(x,t)=e^{i(kx-\omega t)}$ where $\omega=E/\hbar$ and the dispersion relation $k^2=E^2/\hbar^2c^2-m^2c^2/\hbar^2$. 

Using the exact representation \eqref{exactP}, the $q=0$ deformed KG equation can be written as
\begin{eqnarray}
c^2 \Bigl(\frac{1-e^{-2\alpha p_x}}{2\alpha}\Bigr)^2\phi(x,t)=(E^2-m^2c^4)\phi(x,t).\label{kge3}
\end{eqnarray}
Suppose we take the anstaz $\phi(x,t)=e^{i(k' x-\omega t)}$ and choose the undeformed representions of $(p_{x}, E)$ as in \eqref{kge2}, we have 
\begin{eqnarray}
&&\Bigl(\frac{1-e^{-2\alpha\hbar k'}}{2\alpha}\Bigr)^2=\hbar^2 k^2 \Rightarrow\ k'_{\pm}=\frac{-\ln(1\mp 2\alpha\hbar k)}{2\alpha\hbar}\label{wk'}
\end{eqnarray}
and we see that $k'_{\pm}\rightarrow \pm k$ smoothly when $\alpha\rightarrow 0$. Note that wave vector \eqref{wk'} is an exact expression since we have used the exact representation \eqref{rep.P}. 

The wave solution of $q=0$ deformed KG equation is
\begin{eqnarray}
\phi(x,t)&\propto& e^{-i\omega t}(e^{i k'_{+}x}+e^{i k'_{-}x})\nonumber\\
&=& e^{-i\omega t}\Bigl[e^{\frac{-i\ln(1-2\alpha\hbar k)x}{2\alpha\hbar}}+e^{\frac{-i\ln(1+2\alpha\hbar k)x}{2\alpha\hbar}}\Bigr]\nonumber\\
&\approx& e^{-i\omega t}\Bigl[e^{i k x (1 + \alpha \hbar k)}+e^{-i k x (1- \alpha \hbar k)}\Bigr]\label{kgwf}
\end{eqnarray}
where $k$ is the usual wave vector that satisfies the dispersion relation. For the maximum momentum state $k=1/(2\alpha\hbar)$, the wave function \eqref{kgwf} oscillates rapidly. Furthermore, we can define the non-degenerate effective mass $m_{\pm}$ to be proportional to the effective wave vector as
\begin{eqnarray}
\Bigl(\frac{m_{\pm}c}{\hbar}\Bigr)^2 &=& \frac{\omega^2}{c^2}-(k'_{\pm})^2,
\end{eqnarray}
which can be further written as
\begin{eqnarray}
m_{\pm}&=&\frac{\hbar\omega}{c^2}\sqrt{1-\left[\frac{c \ln \bigl(1\mp 2\alpha\sqrt{\hbar^2\omega^2-m^2c^4}/c\bigr)}{2 \alpha \hbar \omega}\right]^2}\nonumber\\
&\approx& m \left[1 \mp m c \left(\frac{\hbar^2 \omega^2}{m^2c^4}-1\right)^{3/2}\alpha\right] + O(\alpha^2).\label{effm}
\end{eqnarray}

We see that both effective mass $m_{\pm}$ smoothly reduce to the ordinary mass when $\alpha\rightarrow 0$. Thus, the $q=0$ deformed KG equation with MCR\eqref{class} describes two particles with effective mass $m_{\pm}$. We constrast this result to the stringy model with minimal length, i.e. Quesne-Tkachuk algebra \cite{quesne} in which the latter consists of one ordinary particle and one Weyl (conformal) ghost. The ghost particle breaks the usual Lorentz symmetry and gives negative contribution to the energy. As a result, the Hamiltonian is not bounded from below in stringy model and one has to introduce indefinite metrics in the field theory sector \cite{moayedi1}.     
  
Similarly, for $q=1$, we have 
\begin{eqnarray}
k'_{\pm}&=&\frac{k}{\pm 1-k \alpha \hbar}\\
\phi(x,t)&\propto& e^{-i\omega t}\Bigl[e^{\frac{ikx}{1-k \alpha \hbar}}+e^{\frac{-ikx}{1+k \alpha \hbar}}\Bigr]\nonumber\\
&\approx& e^{-i\omega t}\Bigl[e^{ ikx (1 + \alpha \hbar k)}+e^{- ikx (1- \alpha \hbar k)}\Bigr].\label{kgwf1}
\end{eqnarray}
The maximum momentum state occurs at $k=1/(\alpha\hbar)$ where the wave function \eqref{kgwf1} oscillates rapidly. Similar to $q=0$, the effective masses are 
\begin{eqnarray}
m_{\pm}&=&\frac{\hbar\omega}{c^2}\sqrt{1-\left[\frac{1-m^2c^4/(\hbar^2\omega^2)}{(\pm 1 -\alpha \sqrt{\hbar^2\omega^2-m^2c^4}/c)^2}\right]}\nonumber\\
&\approx& m \left[1 \mp m c \left(\frac{\hbar^2 \omega^2}{m^2c^4}-1\right)^{3/2}\alpha\right] + O(\alpha^2).\label{effm1}
\end{eqnarray}
To leading order $O(\alpha)$, \eqref{effm1} is the same as the $q=0$ case.

\subsection{MCR and the Real Klein-Gordon equation with scalar and vector linear potential}

Consider the KG equation for a spinless particle of mass $m$ in the presence of a linear scalar $S(x)$ and vector $V(x)$ potential \cite{chargui1}
\begin{eqnarray}
\bigl(c^2 P^2+ [mc^2 + S(X)]^2-[E - V(X)]^2\bigr)\phi=0.\label{MKG1}
\end{eqnarray}
The vector potential is obtained by gauge invariant minimal coupling through $P_{\mu}\rightarrow P_{\mu}-g A_{\mu}$ where $g$ is the real gauge coupling constant of the interaction. We set the spatial component of vector field $\vec{A}=0$ and write the time component as $S(X)=g A_{0}(X)$. This can be viewed as the position dependent rest mass term, i.e. the MIT bag model in quark confinement to avoid Klein paradox \cite{alberto}.   

The potentials are assumed to be linear and take the form of 
\begin{eqnarray}
S(X)&=&\lambda X\ , \hspace{0.5cm}V(X)=\kappa X \label{pots}
\end{eqnarray}
where $(\lambda,\kappa)$ are constants describing the strength of the potentials and have dimension of $Vm^{-1}$. With these potentials, \eqref{MKG1} in momentum space takes the form
\begin{eqnarray}
\left[U^2+\frac{c^2 P^2}{\lambda^2-\kappa^2}-\Bigl(\frac{\lambda mc^2+\kappa E}{\lambda^2-\kappa^2}\Bigr)^2+\frac{m^2c^4-E^2}{\lambda^2-\kappa^2}\right]\phi(P)=0\label{MKG2}
\end{eqnarray}
where we have completed a square and define the scaled position operator as
\begin{eqnarray}
U=X+\frac{\lambda mc^2+\kappa E}{\lambda^2-\kappa^2}. \label{defU}
\end{eqnarray}

For $q=0$, we use the representation $P=p,\ X=i\hbar f(p)\partial_{p}\equiv -2i\alpha\hbar\partial_{\rho}$ where we have performed a change of variable\footnote{In $\rho$ representation, \eqref{MKG2} can be rewritten to a Schrodinger-like equation.} $p=(1-e^{\rho})/(2\alpha)$ which map $p\in(-\infty,\infty)$ to $\rho\in(-\infty,\infty)$. Since $U$ and $X$ are related by a constant shift, we have $[U,P]=[X,P]$. This allows us to choose $U = -2i \hbar \alpha \partial_{\rho}$. With this representation, we have
\begin{eqnarray}
\left[-\frac{\partial^2}{\partial\rho^2}+V_{1}(\rho)\right]\phi(\rho)=\bar{\epsilon}\ \phi(\rho) \label{rhorep1}
\end{eqnarray}
with the definition of dimensionless parameters defined as following
\begin{eqnarray}
\epsilon&=&\frac{m^2c^4-E^2}{4\alpha^2\hbar^2(\lambda^2-\kappa^2)}\ ;\hspace{1.35cm}\theta=\frac{\lambda mc^2+\kappa E}{2\alpha\hbar(\lambda^2-\kappa^2)}\nonumber\\
\delta&=&2\alpha\sqrt{\frac{\hbar(\lambda^2-\kappa^2)^{1/2}}{c}}\ ;\hspace{1cm}\bar{\epsilon}=\theta^2-\epsilon\label{parameters}
\end{eqnarray}
and the effective potential 
\begin{eqnarray}
V_{1}(\rho)&=&\frac{(1-e^{\rho})^2}{\delta^4},\ \hspace{1cm} -\infty<\rho<\infty.
\end{eqnarray}

The potential is similar to Morse potential which has one minimum at $V_{1}(\rho=0)=0$, approaches positive infinity as $\rho\rightarrow \infty$ and approaches $1/\delta^4$ as $\rho\rightarrow -\infty$. Thus, the bound state spectrum has a maximum energy limited by the depth of the well, $\bar{\epsilon}_{max}\leq 1/\delta^4$, which translates to 
\begin{eqnarray}
\bigl(\lambda E_{max}+\kappa mc^2\bigr)^2\leq \frac{c^2(\lambda^2-\kappa^2)}{4\alpha^2};\hspace{0.5cm}|\lambda|>|\kappa|. \label{ebKG}
\end{eqnarray}
For real energy and mass, we see that bound states can occur only when the scalar potential term $S(x)=\lambda X$ is sufficiently stronger than the vector potential $V(x)=\kappa X$, i.e. $(\lambda^2-\kappa^2)>0$ \cite{suRK1, suRK2}. Also, no bound states exist for the case $S(X)=\pm V(X)$. Furthermore, if potential is introduced as the time component of the Lorentz vector ($\lambda=0, \kappa\neq 0$), there is no bound states and only tunneling solutions arise. All these are consistent with the undeformed case \cite{glasser, dominguez} even with the existence of the intrinsic maximum momentum cutoff.   

In the nonrelativistic limit (i.e. $c\rightarrow \infty$), the energy spectrum ($E=E_{\text{nr}}+mc^2$) reads,
\begin{eqnarray}
E_{\text{nr}}\leq \frac{(1-\kappa/\lambda)}{8m\alpha^2}-\frac{mc^2(1+\kappa/\lambda)}{2}.\label{ebKG1}
\end{eqnarray}
For purely scalar confining potential i.e. $(\lambda\neq 0, \ \kappa=0)$ we have 
\begin{eqnarray}
E^2\leq \frac{c^2}{4\alpha^2}\ ,\ \ \text{and} \hspace{0.35cm}E_{\text{nr}}\leq \frac{1}{8m\alpha^2}-\frac{mc^2}{2}.\label{ebKG2}
\end{eqnarray}
From \eqref{ebKG2} it seems to suggest that there is no bound state solution in the non-relativistic limit when $\alpha>(2 m c)^{-1}$. 

\subsection{The $\rho$-Representation and Exact Result}

We proceed to determine the exact bound state energies and wavefunctions. We just outline the main steps\footnote{The procedure to solve equations such as \eqref{rhorep1} is standard \cite{bender}, readers can find details in \cite{spectra}.}. Firstly, an asymptotic analysis identifies the dominant behaviour of square-integrable solutions at the two ends, and then one uses an ansatz for the wavefunction which includes that information to simplify the differential equation. With a further change of variables 
\begin{eqnarray}
\xi&=&{2\over \delta^2} \exp(\rho) \ ;\hspace{0.5cm}0<\xi<\infty \ , \label{repXi}
\end{eqnarray}
which changes the weighted measure to $\int {d \xi \over 2\alpha \xi}$, and defining
\begin{eqnarray}
k  &\equiv &  \sqrt{ {1\over \delta^4} - \bar{\epsilon}} \ > 0 \ ,   \label{kk} \\
\tilde{n} & \equiv & {1 \over \delta^2} \Bigl[ 1-(1-\delta^4 \bar{\epsilon})^{1/2} \Bigr] -{1 \over 2} \ , \label{tiln}
\end{eqnarray}
we obtain 
\begin{eqnarray}
\phi(\xi) &=& e^{-\xi/2} f(\xi) \ \Bigl( {\delta^2 \xi \over 2} \Bigr)^{k},\label{ansatz}
\end{eqnarray}
where $f(\xi)$ is a remaining function that satisfies the reduced equation
\begin{eqnarray}
\xi\frac{\partial^2 f(\xi)}{\partial\xi^2}+\Bigl(2k+1-\xi\Bigr)\frac{\partial f(\xi)}{\partial\xi} + \tilde{n}f(\xi)&=&0.\label{odeXi}
\end{eqnarray} 

The positivity requirement on $k$ comes from the square-integrability condition at $\xi=0$ ($\rho=-\infty$) and it implies an upper bound on the bound state energies identical to what was deduced earlier from the depth of the potential well,
\begin{equation}
\bigl(\lambda E_{max}+\kappa mc^2\bigr)^2\leq \frac{c^2(\lambda^2-\kappa^2)}{4\alpha^2}. \label{ebKG1}
\end{equation} 

Equation \eqref{odeXi} has two singularities, a regular singular point at $\xi=0$ and an essential singularity at $\xi=\infty$. The function  $f(\xi)$  may be expanded in a Frobenius series about $\xi=0$,
\begin{eqnarray}
f(\xi)&=& \xi^s \sum_{j=0}^{\infty} a_{j}\xi^{j} \nonumber \,. 
\end{eqnarray} 
The indicial equation that follows is
\begin{equation}
s (s+ 2k) =0 \, . \nonumber
\end{equation}
The solution with $s=-2k$ is not square-integrable at $\xi=0$ and is discarded. The remaining $s=0$ case leads to an infinite series which must be truncated, otherwise its growth would be exponential and again lead to an unnormalisable solution (at $\xi=\infty$). Truncation leads to the quantisation condition 
\begin{equation}
\tilde{n} =n ,\hspace{0.75cm} (n=0,1,2...) \nonumber
\end{equation}
and thus from (\ref{kk},\ref{tiln}) we obtain 
\begin{eqnarray}
\varepsilon_{n(\alpha)}:= (\lambda E_{n}+ \kappa mc^2) &=& \pm \varepsilon_{n(0)} \sqrt{\left(1- \frac{\alpha^2(\varepsilon_{n(0)})^2}{c^2(\lambda^2-\kappa^2)}\right)}  \, ,  \label{ekg} \\
n  \le  n_{max} &=& \left[{ c \over 4\alpha\hbar\sqrt{\lambda^2-\kappa^2}} - {1 \over 2}\right].
\end{eqnarray}
Here $\varepsilon_{n(0)}=(\lambda^2-\kappa^2)^{3/4}\sqrt{(2n+1)\hbar c}$ is the energy spectrum for the undeformed case which coincides with the result from Ref.\cite{dominguez}. If $n_{max}$ is not an integer, then the largest bound state realized is for an integer less than or equal to the value indicated. In fact $E_n$ in (\ref{ekg}) is a monotonically increasing function of $n$, reaching a stationary point at an integer near $n_{max}$ where  $E=E_{max}$ (\ref{ebKG}) is attained, see Fig.(1). Energy function \eqref{ekg} forbid the complex solution due to the truncation of $n_{max}$ regardless of $\alpha$ parameter. The exact bound states spectrum for $q=0$ has only an $\alpha^2$ correction, which shows the same behaviour as the deformed harmonic oscillator (DHO).

Next, the reduced Schrodinger equation is
\begin{eqnarray}
\xi\frac{\partial^2 f(\xi)}{\partial\xi^2}+\Bigl(2k_n +1-\xi\Bigr)\frac{\partial f(\xi)}{\partial\xi}+ nf(\xi)&=&0; \hspace{0.25cm}n\ni Z^{+} \label{wavef1}
\end{eqnarray}
with solutions given by  associated Laguerre polynomials $f(\xi)=L_{n}^{2k_n}(\xi)$. The normalized wave functions are \cite{bender}
\begin{eqnarray}
\Psi_{n}^{2k_{n}}(\xi)&=&\sqrt{\frac{ 4 \alpha k_n n!}{\Gamma(2k_n+n+1)}} \exp(-\xi/2)\xi^{k_n} L_{n}^{2k_n}(\xi) \, \label{wavef2}
\end{eqnarray}    
with the following normalization condition,
\begin{eqnarray}
\int_{0}^{\infty} \frac{d\xi}{2\alpha\xi}\Bigl|\Psi_{n}^{2k_{n}}(\xi)\Bigr|^2=1.
\end{eqnarray}
To illustrate the behaviour of the eigenfunctions, we plot an example $\Psi_{n}^{2k_{n}}(\xi)$ in Fig.(2) for $n=10$. For small $\alpha$ and $n$, the wavefunction does not feel the maximum momentum. In Fig.(2), it shows the probability density for the $n=10$ state which is already slightly asymmetrical with higher peak shifted closer to $p=p_{max}$ (corresponds to $\xi=0$).

\section{MCR and the Dirac equation with scalar-like and vector-like linear potential}

Consider the stationary Dirac equation describing a spin-$\frac{1}{2}$ particle of mass $m$ subjected to a scalar potential $S(X)$ and vector potential $V(X)$ as in KG equation \cite{chargui2, pedramDE},
\begin{eqnarray}
(E-V(X)) \psi&=& \bigl[c\vec{\alpha}\cdot\vec{P}+\beta(mc^2+S(X))\bigr]\psi, \label{dirac1}
\end{eqnarray}
where $\psi$ is the Dirac spinors and $(\vec{\alpha},\beta)$ the Dirac Matrices. In (1+1)-dimension, $\psi$ is represented by two components spinor and we can choose the anti-commuting Dirac matrices $(\vec{\alpha},\beta)$ as
\begin{eqnarray}
\alpha&=&\sigma_{y}=\begin{pmatrix}
 0 & -i \\
 i & 0
\end{pmatrix}\ ;\hspace{1cm}
\beta=\sigma_{z}=\begin{pmatrix}
 1 & 0 \\
 0 & -1
\end{pmatrix}.
\end{eqnarray}

Suppose we use the ansatz for the spinor $\psi$ as
\begin{eqnarray}
\psi&=&\bigl[c\alpha P+\beta(mc^2+S(X))+(E-V(X))\bigr]\phi \label{ansatz1}
\end{eqnarray}
where \eqref{ansatz1} can be viewed as the ``transformation'' in the spinor space and $\phi$ is a two components function satisfying the following,
\begin{eqnarray}
\Bigl[c^2P^2+(mc^2+\lambda X)^2-(E-\kappa X)^2-c[P,(\beta\lambda+\kappa) X]\alpha\Bigr]\phi&=&0
\end{eqnarray}
where we assume the linear potential \eqref{pots}. After completing square and substituting the commmutator $[P,X]$ and Dirac matrices, it leads to 
\begin{eqnarray}
\left[U^2+\frac{c^2 P^2}{\lambda^2-\kappa^2}-\Bigl(\frac{\lambda mc^2+\kappa E}{\lambda^2-\kappa^2}\Bigr)^2+\frac{m^2c^4-E^2}{\lambda^2-\kappa^2}+\frac{\hbar c (1-2\alpha P)}{\sqrt{\lambda^2-\kappa^2}}M\right]\phi(P)=0\label{mde1}
\end{eqnarray}
in which we have defined the scaled position operator as in \eqref{defU}. Here $M$ is the matrix defined as
\begin{eqnarray}
M:= \frac{1}{\sqrt{\lambda^2-\kappa^2}}\begin{pmatrix}
 0 & \lambda+\kappa \\
\lambda-\kappa & 0
\end{pmatrix}
\end{eqnarray}
with eigenvalues of $\eta=\pm 1$ and two components eigenvector $R_{\eta}$, i.e. $M R_{\eta}=\eta R_{\eta}$. We assume the ansatz $\phi(P)=R_{\eta}\varphi_{\eta}(p)$ where $\varphi_{\eta}(p)$ is a function of momentum characterized by $\eta=\pm 1$. 

With the new position and momentum variable $U=-2i\hbar\alpha\partial_{\rho},\ p=(1-e^{\rho})/(2\alpha)$, we have
\begin{eqnarray}
\left[-\frac{\partial^2}{\partial\rho^2}+V_{2}(\rho)\right]\varphi_{\eta}(\rho)=\bar{\epsilon}\ \varphi_{\eta}(\rho) \label{rhorep2}
\end{eqnarray}
with the potential 
\begin{eqnarray}
V_{2}(\rho)&=&\frac{(1-e^{\rho})^2}{\delta^4}+\frac{\eta e^{\rho}}{\delta^2},\ \hspace{0.75cm} -\infty<\rho<\infty
\end{eqnarray}
and $\bar{\epsilon}$ is defined in \eqref{parameters}. In contrast to potential $V_{1}(\rho)$ in KG equation, $V_{2}(\rho)$ has a shifted minimum $V_{2}(\rho)\bigl|_{\text{min}} = \eta (1/\delta^2 - \eta/4)$ that happens at $\rho = \ln(1-\frac{\eta\delta^2}{2})$. We see that the minimum of $V_{2}(\rho)$ is shifed up (or down) with regards to $V_{1}(\rho)\bigl|_{\text{min}}=0$ for $\eta=1$ (or $\eta=-1$) case (see Fig.(3)). Furthermore, similar to $V_{1}(\rho)$, $V_{2}(\rho)$ saturates to $1/\delta^4$ when $\rho\rightarrow -\infty$ regardless of $\eta$. Thus, the energy of the MCR-inspired generalized Dirac equation with linear potential is bounded from above which is similar to the Klein-Gordon case \eqref{ebKG}, i.e. $(\lambda E_{max}+\kappa mc^2)^2\leq c^2(\lambda^2-\kappa^2)/(4\alpha^2)$.     

Since \eqref{rhorep2} has same structure as \eqref{rhorep1} with new effective potential $V_{2}(\rho)$, one can use the same method as in Sect.(3) to determine the exact bound state energies and wavefunctions. After a change of variable and apply the ansatz solution $\phi(\xi) = e^{-\xi/2} f(\xi)\ (\delta^2 \xi/2)^{k}$, we obtain from \eqref{odeXi} 
\begin{eqnarray}
\xi\frac{\partial^2 f(\xi)}{\partial\xi^2}+\Bigl(2k+1-\xi\Bigr)\frac{\partial f(\xi)}{\partial\xi} + n' f(\xi)&=&0. \nonumber
\end{eqnarray} 
Here, the definition of $k$ is similar to \eqref{kk} but $n'$ is modified to
\begin{eqnarray}
n' & \equiv & {1 \over \delta^2} \Bigl[1-\frac{\eta\delta^2}{2}-(1-\delta^4 \bar{\epsilon})^{1/2} \Bigr] -{1 \over 2}. \label{n'}
\end{eqnarray}
The square-integrability condition of wave function at $\xi=0\ \text{or}\ (\rho=-\infty)$ requires $k$ value to be positive definite and hence predicts the same maximum energy as in KG case, $\bar{\epsilon}=1/\delta^4$. Next, to avoid unnormalisable wave function at $\xi=\infty$, $n'$ has to be positive integers. Since we have the same maximum energy cutoff, the maximum number of bound states is given by
\begin{eqnarray}
n'&\leq & n'_{max}=\left[\frac{c}{4\alpha\hbar\sqrt{\lambda^2-\kappa^2}} -\frac{\eta}{2}- \frac{1}{2}\right].
\end{eqnarray} 

From \eqref{n'} we obtain energy quantization condition as
\begin{eqnarray}
\bar{\epsilon}&=&\frac{1-\bigl(1-\frac{\eta\delta^2}{2}-(n'+\frac{1}{2})\delta^2\bigr)^2}{\delta^4}=2n(1/\delta^2-n/2)\label{ede}
\end{eqnarray}
where we have defined $n\equiv n'+1/2+\eta/2$. The $n$ value can take all nonzero positive integers up to $n_{max}$ for both $\eta=\pm 1$, with $n=0$ is only possible for $\eta=-1$ case. It is interesting to see that for non-zero principle quantum number, i.e. $n\neq 0$, we have two possible pairs of $(\eta,n')$ that correspond to the doubly degeneracies in the energy. They are $(-1,n)\ \text{or}\ (1,n-1)$. This result is universal to MCR's models including those with nonzero minimal length \cite{chargui2, pedramDE} and those with intrinsic maximum momentum. The energy of the MCR-inspired Dirac equation with vector and scalar potential has similar form to \eqref{ekg}
\begin{eqnarray}
\varepsilon_{n(\alpha)}:= (\lambda E_{n}+ \kappa mc^2) &=& \pm \varepsilon_{n(0)} \sqrt{\left(1- \frac{\alpha^2(\varepsilon_{n(0)})^2}{c^2(\lambda^2-\kappa^2)}\right)}  \, ,  \label{ede1}
\end{eqnarray}
where $\varepsilon_{n(0)}=(\lambda^2-\kappa^2)^{3/4}\sqrt{2n\hbar c}$\ is the energy expression for the undeformed case which coincides exactly the result from literature \cite{suRK1}. Here $n=1,2...$ for $\eta=+1$ and $n=0,1,2,3...$ for $\eta=-1$ case. If one consider to use $n'$ to label the principle quantum number, the energy spectrum reads as,
\begin{eqnarray}
\lambda E^{(+)}_{n'}+\kappa mc^2&=&\pm 2\alpha\hbar(\lambda^2-\kappa^2)\sqrt{2(n'+1)/\delta^2 - (n'+1)^2}\\
\lambda E^{(-)}_{n'}+\kappa mc^2&=&\pm 2\alpha\hbar(\lambda^2-\kappa^2)\sqrt{2n'/\delta^2 - (n')^2}.
\end{eqnarray}
Also, the bound state wavefunctions are exactly given by \eqref{wavef2} with the replacement $n\rightarrow n'$.

The difference in maximum number of states for both $\eta=\pm 1$ case is $\tilde{n}_{max}^{(-)}-\tilde{n}_{max}^{(+)}=1$. This is due to the difference in the depth of effective potential $V_{2}(\rho)$ for both $\eta=\pm 1$. In fact, we see that $\eta=-1$ case captures one additional state that reads $(\eta,n')=(-1,0)$ or equivalently $n=0$ (See Fig.(4)). This scenario is consistent with the potential in Fig.(1) in which $V_{2}^{(-1)}$ has real root at $\bar{\epsilon}=0$ while $V_{2}^{(+1)}$ is shifted up from the
$\bar{\epsilon}=0$ line since $1/\delta^2-1/4>0$ for small deformed parameter $\alpha$. We emphasize that $n=0$ state is absent in undeformed case since it does not belong to bound state solution \cite{chargui2}.    

To summarize, our result is similar in several aspects to the model in which there is a nonvanishing minimal length. For instance, there is an appearance of an additional energy level corresponding to the quantum number $n=0$ which is not a bound state solution for undeformed case. Nevertheless, the bound states energy spectrum are different and bounded from above for MCR case. MCR shifts lower an undeformed positive energy level and raises an undeformed negative energy level. These are constrasting to the model with minimal length. Furthermore, there is a maximum number of bound states $n_{max}$ which is absent in the GUP case.     

\section{MCR and the Dirac Oscillator}

The Dirac oscillator is motivated in relativistic quantum mechanics to obtain a potential that is linear in both the momentum and spatial coordinates, and thus exhibits nontrivial equation of motion. It was shown to be `square root' of linear harmonic oscillator plus a strong spin-orbit coupling term \cite{martinez}. One applies the non-minimal substitution in the free dirac Hamiltonian with $\vec{P}\rightarrow \vec{P}-i m\omega \beta\vec{r}$, where $m$ is the mass of the dirac particle and $\omega$ is the oscillator frequency. The Dirac oscillator equation is
\begin{eqnarray}
i\hbar\frac{\partial}{\partial t}\psi&=&H\psi=\left[c\vec{\alpha}\cdot\bigl(\vec{P}-i m\omega \beta\vec{r}\bigr)+mc^2\beta\right]\psi
\end{eqnarray}
where $\psi=(\phi,\chi)^T$ is the Dirac spinor and $(\vec{\alpha},\beta)$ are the Dirac matrices. The Hamiltonian is manifestly Hermitian by the property of the Dirac $\alpha$ matrix. 

For stationary solution in one-dimension, we have
\begin{eqnarray}
E \psi&=&\bigl[c\alpha\bigl(P-i m\omega\beta X\bigr)+mc^2\beta\bigr]\psi.  
\end{eqnarray}
After choosing the representation $\alpha=\sigma_{x}=\begin{pmatrix}
 0 & 1 \\
 1 & 0
\end{pmatrix}$ 
and $\beta=\sigma_{z}=\begin{pmatrix}
 1 & 0 \\
 0 & -1
\end{pmatrix}$, we have
\begin{eqnarray}
c(P+im\omega X)\chi&=&(E-mc^2)\phi \label{doe1}\\
c(P-im\omega X)\phi&=&(E+mc^2)\chi. \label{doe2}
\end{eqnarray}
From \eqref{doe2}, we can express $\chi=\frac{c(P-im\omega X)}{E+mc^2}\phi$ and substitute into \eqref{doe1} to obtain
\begin{eqnarray}
\Bigl(\frac{E^2}{c^2}-m^2 c^2\Bigr)\phi&=&\bigl(P^2+m^2\omega^2 X^2+i m \omega [X,P]\bigr)\phi. \label{doe3}
\end{eqnarray}
Using $[X,P]=i\hbar f(P)$, where $f(P)=1-2\alpha P+q\alpha^2 P^2$ and the representation \eqref{rep.P}-\eqref{rep.X}, we can rewrite \eqref{doe3} as
\begin{eqnarray}
&\phantom{|}&\Bigl[(1-q m\hbar\omega\alpha^2)p^2+(2m\hbar\omega\alpha) p-(m\hbar\omega f(p)\partial_{p})^2\Bigr]\phi\nonumber\\
&&=\Bigl(E^2/c^2-m^2 c^2+m\hbar\omega\Bigr)\phi
\end{eqnarray}
for general $q$. For the case $q=0$, after dividing both side by $(2m\hbar\omega\alpha)^2$, we have
\begin{eqnarray}
\left[\Bigl(\frac{p}{2m\hbar\omega\alpha}+\frac{1}{2}\Bigr)^2-\Bigl(\frac{f(p)\partial_{p}}{2\alpha}\Bigr)^2\right]\phi&=&\left[\frac{E^2-m^2 c^4}{(2mc\hbar\omega\alpha)^2}+\frac{1}{4m\hbar\omega\alpha^2}+\frac{1}{4}\right]\phi.\nonumber\\
\end{eqnarray}
Similar to previous section, by performing a change to $\rho$ variable, i.e. $\partial_{\rho}\equiv-f(p)\partial_{p}/(2\alpha)$ we obtain the Schrodinger-like equation, 
\begin{eqnarray}
\left[-\frac{\partial^2}{\partial\rho^2}+V_{3}(\rho)\right]\phi(\rho)=\bar{\epsilon}_{1}\ \phi(\rho) \label{rhorep3}
\end{eqnarray}
with the effective energy parameter
\begin{eqnarray}
\bar{\epsilon}_{1}&=&\frac{E^2-m^2 c^4}{(2mc\hbar\omega\alpha)^2}+\frac{1}{4m\hbar\omega\alpha^2}+\frac{1}{4}
\end{eqnarray}
and the potential 
\begin{eqnarray}
V_{3}(\rho)&=&\Bigl(\frac{1-e^{\rho}}{\delta_{1}^2}+\frac{1}{2}\Bigr)^2;\ \hspace{1cm} -\infty<\rho<\infty,
\end{eqnarray}
where $\delta_{1}=2\sqrt{m\hbar\omega}\alpha$. The potential, similar to Morse potential in Fig.(1), has one minimum at $V_{3}(\rho=\ln(1+\delta_{1}^2/2))=0$, approaches positive infinity as $\rho\rightarrow \infty$ and approaches $(\frac{1}{\delta_{1}^2}+\frac{1}{2})^2$ as $\rho\rightarrow -\infty$. Thus, the bound state spectrum has a maximum energy limited by the depth of the well, $\bar{\epsilon}_{1,max}\leq (\frac{1}{\delta_{1}^2}+\frac{1}{2})^2$, which translates to 
\begin{eqnarray}
E_{max}^2\leq m^2c^4+\frac{c^2}{4\alpha^2}. \label{ebde}
\end{eqnarray}
In the nonrelativistic limit ($c\rightarrow\infty$), the energy spectrum ($E=E_{\text{nr}}+mc^2$) reads,
\begin{eqnarray}
E_{\text{nr}}\leq \frac{1}{8m\alpha^2}.
\end{eqnarray}
This is the same as the upper energy bound of non-relativistic MCR-deformed DHO\cite{spectra}. It indicates that there exist a maximum energy/momentum in which there are no bound states. 

Introducting a variable $\zeta^2=\frac{2\delta_{1}^2}{2+\delta_{1}^2}$, we can rewrite $V_{3}(\rho)$ as similar form to $V_{1}(\rho)$ in Sect.(3)
\begin{eqnarray}
V_{3}(\rho)&=&\frac{(1-e^{\rho+\ln(\zeta^2/\delta_{1}^2)})^2}{\zeta^4} \equiv \frac{(1-e^{\bar{\rho}})^2}{\zeta^4},\ \hspace{0.35cm} -\infty<\bar{\rho}<\infty.
\end{eqnarray}
where we have performed a constant shift $\rho\rightarrow \bar{\rho}=\rho+\ln(\zeta^2/\delta_{1}^2)$. In $\bar{\rho}$ variable, we can apply the same formalism in Sect.(3) to write down the Schrodinger-like equation and it's ansatz solution,
\begin{eqnarray}
\psi(\xi) = e^{-\xi/2} f(\xi)\ (\delta^2 \xi/2)^{k},\hspace{0.5cm} 0<\xi<\infty
\end{eqnarray}
in which the reduced Schrodinger equation is again given by \eqref{odeXi}
\begin{eqnarray}
\xi\frac{\partial^2 f(\xi)}{\partial\xi^2}+\Bigl(2k+1-\xi\Bigr)\frac{\partial f(\xi)}{\partial\xi} + \tilde{n}f(\xi)&=&0. \nonumber
\end{eqnarray}
The definition of $(k,\tilde{n})$ are similar to \eqref{kk}-\eqref{tiln} but have to replace the $(\delta \rightarrow\zeta, \bar{\epsilon}\rightarrow\bar{\epsilon}_{1})$
\begin{eqnarray}
k&=&\sqrt{\frac{1}{\zeta^4}-\bar{\epsilon}_{1}}>0\ ;\\
\tilde{n}&=&\frac{1}{\zeta^2}\Bigl[1-(1-\zeta^4\bar{\epsilon}_{1})^{1/2} \Bigr] -\frac{1}{2}.\label{tiln2}
\end{eqnarray}
Similar to the previous sections, the square-integrability condition of wave function at both $\xi=0$ and $\xi=-\infty$ requires $k$ value to be positive definite and predicts the maximum energy, $\bar{\epsilon}_{1}\leq 1/\zeta^4$. Also, $\tilde{n}$ has to be positive integers, i.e. $\tilde{n}=n$ and bounded from above,
\begin{eqnarray}
n&\leq &n_{max}=\frac{1}{\delta^2_{1}}=\left[\frac{1}{4 m \hbar \omega \alpha^2}\right].
\end{eqnarray} 

From \eqref{tiln2} we obtain energy quantization condition as
\begin{eqnarray}
\bar{\epsilon}_{1}&=&\frac{1-\bigl(1-(n+\frac{1}{2})\zeta^2\bigr)^2}{\zeta^4}=\Bigl(n+\frac{1}{2}\Bigr)\left[\frac{2-\bigl(n+\frac{1}{2}\bigr)\zeta^2}{\zeta^2}\right].\label{edo}
\end{eqnarray}
Thus the energy of MCR-inspired Dirac oscillator is exactly given by
\begin{eqnarray}
E_{n(\alpha)}&=&\pm E_{n(0)}\sqrt{\left[1- \frac{\alpha^2\bigl(E_{n(0)}-m^2c^4/E_{n(0)}\bigr)^2}{c^2}\right]}  \, ,  \label{edo1}
\end{eqnarray}
where $E^2_{n(0)}=m^2c^4\bigl(1+2n\hbar\omega/(mc^2)\bigr)$ is the energy expression for the undeformed case which coincides exactly the result from literature \cite{martinez}-\cite{mosh}. The nonrelativistic limit of \eqref{edo1} is obtained as 
\begin{eqnarray}
E^{nr}_{n(\alpha)}&=&\pm E^{nr}_{n(0)}\bigl[1-2m \alpha^2 E^{nr}_{n(0)}\bigr]\, ,  \label{edo2}
\end{eqnarray}
and we consistently reproduce the non-relativistic DHO \cite{spectra}. In addition, the bound state wavefunctions are given by \eqref{wavef2} with the appropriate replacement in $(\delta,\epsilon)$ variables.
 
The expectation values of the position/momentum operator in the state \eqref{wavef2} can be evaluated as
\begin{eqnarray}
\langle X\rangle&=&0\\
\langle X^2\rangle&=&\Bigl(n+\frac{1}{2}\Bigr)\frac{\hbar}{m \omega}\Bigl[1- \frac{n}{n_{max}}\Bigr]\\
\langle P\rangle&=&\Bigl(\frac{n}{n_{max}}\Bigr)\ \sqrt{m \hbar \omega n_{max}}\\
\langle P^2\rangle&=&m \hbar \omega \Bigl(n+\Bigl[\frac{n_{max}-n}{2n_{max}}\Bigr]\Bigr)
\end{eqnarray} 
and thus we have the uncertainties, i.e. $(\Delta O)^2:=\langle O^2\rangle-\langle O\rangle^2$
\begin{eqnarray}
(\Delta X)^2&=&\Bigl(n+\frac{1}{2}\Bigr)\frac{\hbar}{m \omega}\Bigl[1- \frac{n}{n_{max}}\Bigr]\\
(\Delta P)^2&=&\Bigl(n+\frac{1}{2}\Bigr) m \hbar \omega \Bigl[1- \frac{n}{n_{max}}\Bigr]\\
\Rightarrow (\Delta X)(\Delta P)&=&\Bigl(n+\frac{1}{2}\Bigr)\hbar\Bigl[\frac{n_{max}-n}{n_{max}}\Bigr].
\end{eqnarray} 
We see that at the maximum energy state $n=n_{max}$, the uncertainties $(\Delta X)$ and $(\Delta P)$ vanish and hence leads to classical phase as similar to non-relativistic DHO.

\section{Conclusion}

We have studied relativistic wave equation under the scheme of one-dimensional deformed Quantum Mechanics with intrinsic maximum momentum, but without exitence of a minimal length, i.e. $q\leq 1$. Modified dispersion relation is studied and we have the superluminal character of photon propagation in MCR\eqref{class} which is consistent with varying speed of light (VSL) phenomena. Solution of free Klein-Gordon equation is obtained and at maximum momentum, the wavefunction oscillates rapidly. We have two particles solution with effective masses $m_{\pm}$ that change smoothly to ordinary mass in the undeformed limit.

Using momentum space representation, we have solved exactly the Klein-Gordon/Dirac equation with linear confining scalar/vector potentials and the deformed Dirac oscillator. We have found the exact energy spectrum, eigenfunctions and have shown that the correct non-relativistic, undeformed limit are reproduced. Similar to non-relativistic DHO \cite{spectra}, the energies of these systems are bounded from above due to the intrinsic maximum momentum cutoff. There is a maximum allowed bound state solution depending on the deformed parameter $\alpha$. For Dirac equation with linear potentials, we have obtained an additional energy level with quantum number $n=0$ due to the maximum momentum. This effect cannot be distinguished from the GUP case with minimal length \cite{chargui2}. For Dirac oscillator, it's non-relativistic limit is consistent with the deformed harmonic oscillator. The study in this paper may have potential applications in quantum confinement and quantum gravity phenomenology.

\section{Acknowledgment}
C.L Ching would like to thank Dr. Parwani for great and stimulating discussions.

\section*{Figures}
 
\begin{figure}[ht]
\begin{center}
\epsfig{file=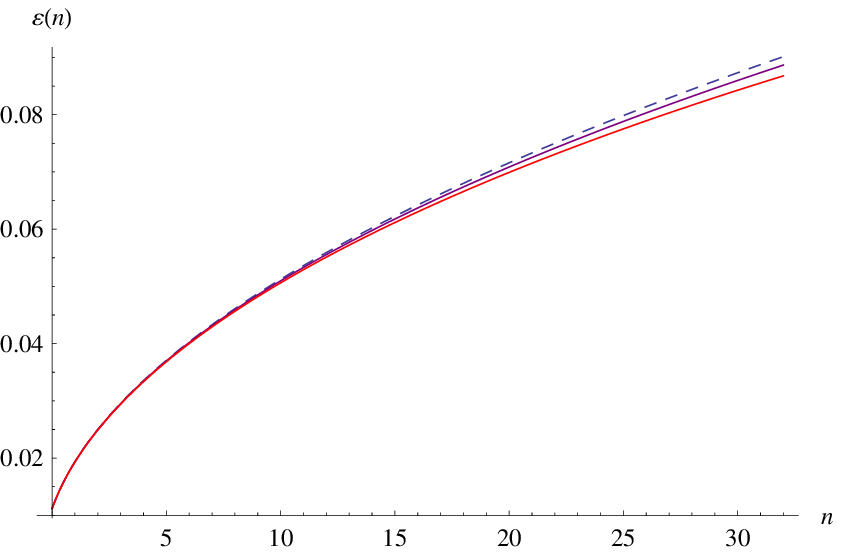, width=12cm}
\caption{The effective potential for $V_{2}^{\eta}(\rho)$. Note that the potential is saturated to $\frac{1}{\delta^4}$ when $\rho\rightarrow -\infty$ as similar to $V_{1}(\rho)$. Dashed (solid) line refers to $\eta=-1\ (\eta=+1)$ case. We set $\delta=0.5$.}
\label{e1}
\end{center}
\end{figure}

\begin{figure}
  \begin{center}
   \epsfig{file=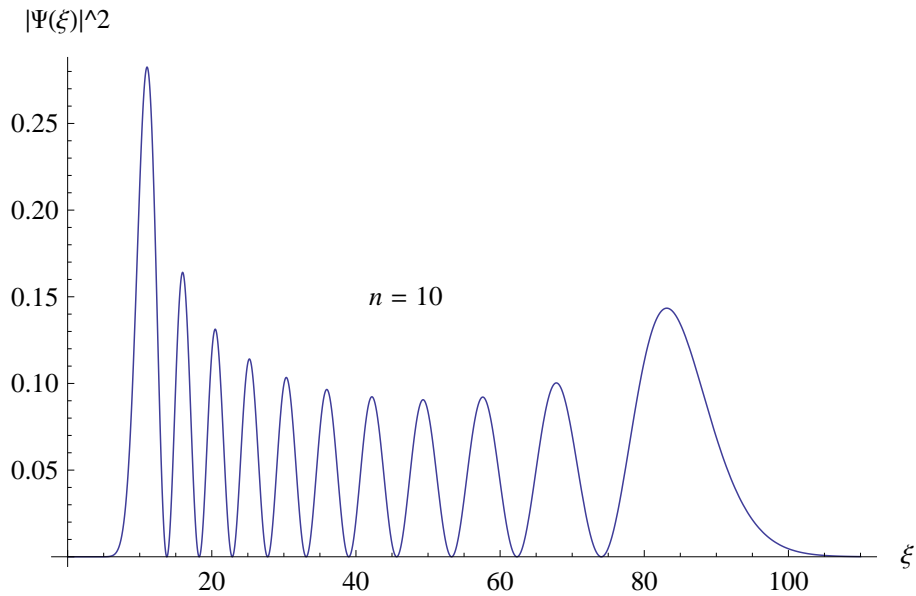, width=12cm}
    \caption{The characteristic $\varepsilon_{n}$ energy spectrum for Klein-Gordon equation. We set $\lambda=0.05, \kappa=0; \hbar=c=1$. Dashed line refers to undeformed case. Purple/Red lines correspond to $(\alpha=0.1; 0.15)$. Note $n_{\text{max}}|_{\alpha=0.1}=49$ and $n_{\text{max}}|_{\alpha=0.15}=32$.}
    \label{e2}
  \end{center}
\end{figure}

\begin{figure}
  \begin{center}
   \epsfig{file=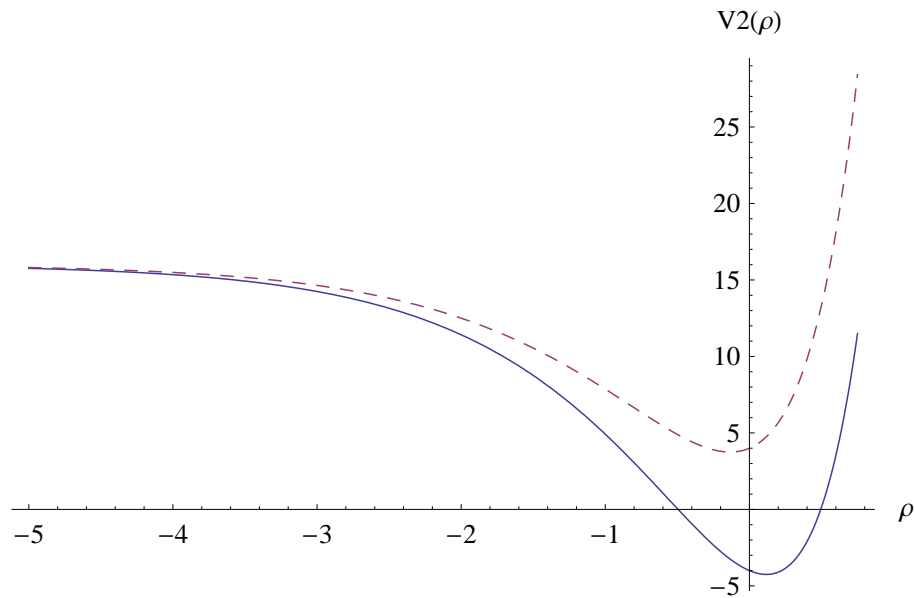, width=12cm}
    \caption{The probability density for $q=0, n=10$. The horizontal axis is $\xi =2(1-2\alpha P)/\delta^2$ and so $\xi=0$ corresponds to $p=p_{max}=1/(2 \alpha)$ while $\xi =\infty$ corresponds to $p= -\infty$. We set $\alpha=0.1; \lambda=1, \kappa=0; \hbar=c=1$.}
    \label{e2}
  \end{center}
\end{figure}

\begin{figure}
  \begin{center}
   \epsfig{file=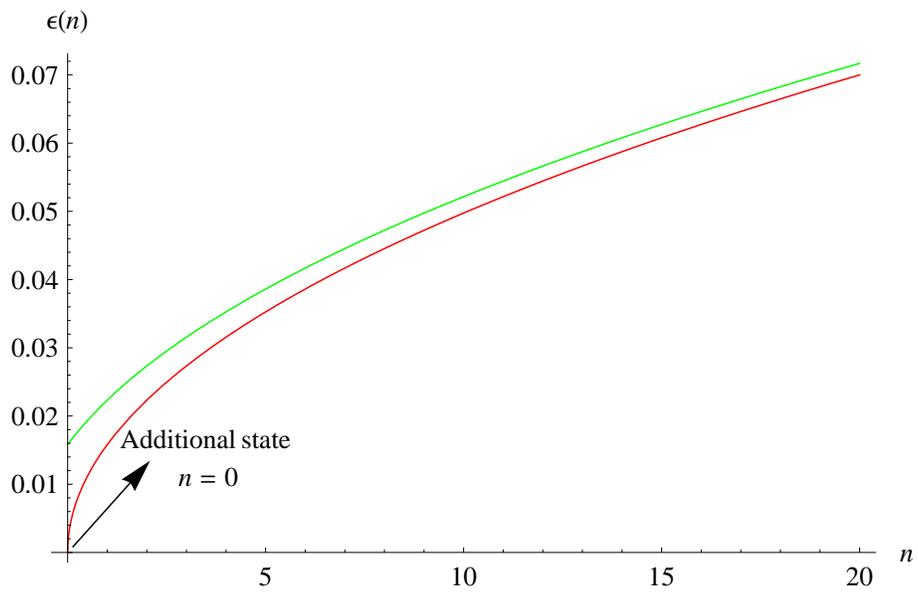, width=12cm}
    \caption{The characteristic $\varepsilon_{n}$ energy spectrum for Dirac equation. We set $\alpha=0.1,\lambda=0.05, \kappa=0; \hbar=c=1$. Green/Red lines correspond to $\eta=1 ; (\eta=-1)$. Note there is an additional state for $\eta=-1$ at $n'=n=0$.}
    \label{e2}
  \end{center}
\end{figure}

\end{document}